\documentclass[11pt,twoside]{article}
\usepackage{graphicx}
\usepackage{amsmath}
\usepackage{amssymb}
\usepackage{mathrsfs}
\usepackage{setspace}

\begin{document}

\thispagestyle{plain}

\newcommand{\be}{\begin{equation}}
\newcommand{\ee}{\end{equation}}
\newcommand{\ds}{\displaystyle}
\newcommand{\bdm}{\begin{displaymath}}
\newcommand{\edm}{\end{displaymath}}
\newcommand{\bea}{\begin{eqnarray}}
\newcommand{\eea}{\end{eqnarray}}
\newcommand{\mbf}{\mathbf}
\newcommand{\rmd}{\mathrm{d}}
\newcommand{\rme}{\mathrm{e}}
\newcommand{\bm}{\boldsymbol}
\newcommand{\vpint}{-\!\!\!\!\!\!\int}

\begin{center} {\Large \bf
\begin{tabular}{c}
Normalized non-redundant vector tomographic \\
portraits of spin states
\end{tabular}
 } \end{center}


\begin{center} {\bf Ya. A. Korennoy}\end{center}


\begin{center}
{\it P. N. Lebedev Physical Institute of the Russian Academy of Science, \\ 
                       Leninskii prospect 53, Moscow 119991, Russia }
\end{center}

\begin{abstract}\noindent
Non-redundant and normalized  four-component  vector  tomographic portrait fully describing the  states of spin $1/2$
quantum particles was introduced. Dequantizer and quantizer for such portrait were found, and generalization to the case 
of spin $(2^N-1)/2$ was done ($N$ is a natural number).
It was shown that such a portrait is completely defined by a thriple 
of non-complanar vectors with the lengthes equal or less then unity.
A clear geometric interpretation of the choice of parameters 
for finding normalized dequantizers and quantizers is presented 
and numerical examples of such dequantizers and quantizers for 
spin $ 1/2 $ are given.
\end{abstract}


\section{Introduction}
As is known, in the traditional formulation of quantum mechanics
the pure states of a quantum particle with spin $s$ are described 
by $(2s+1)$-component complex spinors $(\psi_1,\psi_2,...\psi_{2s+1})$. 
Mixed states of such particles are represented by $(2s+1)\times(2s+1)$-density matrices
whose off-diagonal elements, in general, are also complex. 
On the other hand, the tomographic approach 
(see \cite{IbortPhysScr,MankoMankoFoundPhys2009,LvovRayRevModPhys})
makes it possible to portray the states of quantum systems 
by real nonnegative quantities.
In \cite{DodonovManko1997} the tomographic distribution for rotated spin variables
was constructed, but the method suggested there is inconvenient because of redundant
data containing in the tomogram.

The tomographic description is closely connected with the state reconstruction problem,
to the solution of which for the spin states the following papers were devoted:
\cite{NewtonYoung1968,Weigert1992,Weigert1998,AW12,AW13,AW14,AW15,HeissWeigert2000}.
The scope of this problem is finding of the transformation procedure 
of recovering of the density matrix from the set of expectation values 
of observables constituting a quorum.
In the state reconstruction problem the superfluous amount of data 
for obtaining of the density matrix is possible not to consider as deficiency.  
Often  the extra data can enable
to carry out more exact accounts.
On the contrary, in the tomographic formulation of quantum mechanics
the redundant data in the tomogram for application of the inverse map
is an essential inconvenience.

The tomographic description of systems with spin was also evolved  in 
\cite{MankoMarmo2004,FilippovManko2010} and in other papers.
In \cite{Korarticle9,Korarticle14} it was introduced the positive non-redundant
vector tomographic portrait fully describing the states of quantum particles
including both spatial and spin information.
In the case of consideration of only spin subspace
the essence of this approach is based on the inverse problem
studied in \cite{Weigert1998,AW13,AW14}.
But we will follow the dequantizer -- quantizer terminology
used in \cite{Korarticle9,Korarticle14}, where 
the states of quantum particles 
with spin $s$ are described as the $(2s+1)^2$-component vector tomographic portrait
$\mbf w=\left(w_1,w_2,...,,w_{(2s+1)^2}\right)$,
\be			\label{defw}
\mbf w=\mathrm{Tr}\{\hat\rho\, \hat{\bm{\mathcal{U}}}\},
\ee
and $\hat{\bm{\mathcal{U}}}=\left(\hat{\mathcal{U}}_1,\hat{\mathcal{U}}_2,...,
\hat{\mathcal{U}}_{(2s+1)^2}\right)$ is $(2s+1)^2$-component  dequantizer vector
with components $\hat{\mathcal{U}}_k$ that are projectors onto
the $(2s+1)^2$ pure and/or mixed spin states, which are  chosen so that the matrix 
of linear transformation $\hat\rho \to \mbf w$ would be reversible.

To find the inverse transformation of (\ref{defw}) 
the $(2s+1)^2$-component quantizer vector $\hat{\bm{\mathcal{D}}}$ 
is used, whose components 
$\hat{\mathcal{D}}_1$, $\hat{\mathcal{D}}_2$, ..., $\hat{\mathcal{D}}_{(2s+1)^2}$
are the $(2s+1)\times(2s+1)$-matrixes satisfying the conditions
\be			\label{ortcondition}
\mathrm{Tr}\left\{\hat{\mathcal{U}}_j\hat{\mathcal{D}}_{j'}\right\}=
\sum_{k,l=1}^{2s+1}\mathcal{U}_{j(kl)}\mathcal{D}_{j'(lk)}=\delta_{jj'},~~~~
\sum_{j=1}^{(2s+1)^2}\mathcal{U}_{j(kl)}\mathcal{D}_{j(l'k')}=
\delta_{kk'}\delta_{ll'}.
\ee
Here  letters  $j,j'=\overline{1,(2s+1)^2}$ are the  indexes corresponding 
to the numbers of the components of the tomographic vector $\mbf w$, 
and letters in parentheses $(kl)$ or $(k'l')$ are the spin indexes 
of $(2s+1)\times(2s+1)$-matrices.

Using the quantizer $\hat{\bm{\mathcal{D}}}$ the inverse transformation of (\ref{defw})
is written as the scalar product of two vectors
\be			\label{invmap}
\hat\rho=\mbf w\hat{\bm{\mathcal{D}}}.
\ee
The components of the vector $\hat{\bm{\mathcal{U}}}$ or $\hat{\bm{\mathcal{D}}}$
form the basis in the space of $(2s+1)\times(2s+1)$-matrices,
and any $(2s+1)\times(2s+1)$-density matrix is a convex sum 
of the  components of $\hat{\bm{\mathcal{D}}}$.

It follows from (\ref{invmap}) that the normalization condition of $\hat\rho$
can be written in terms of $\mbf w$:
\be			\label{normRhofrom_w}
\mathrm{Tr}\,\hat\rho=\left(\mathrm{Tr}\,\hat{\bm{\mathcal{D}}}\right)\mbf w.
\ee

The components of $\mbf w$ satisfy the conditions $0\leq w_i\leq 1$.
As for the normalization of the vector $\mbf w$, then its existence depends 
on the choice of the projectors $\{\hat{\cal U}_k\}_{k=\overline{1,(2s+1)^2}}$.
It is obvious from (\ref{defw}) that
\be			\label{sumw}
\sum_{j=1}^{(2s+1)^2}w_j=\sum_{j=1}^{(2s+1)^2}\mathrm{Tr}\{\hat\rho\,\hat{\mathcal{U}}_j\}=
\mathrm{Tr}\left\{\hat\rho\,\sum_{j=1}^{(2s+1)^2}\hat{\mathcal{U}}_j\right\}.
\ee
Therefore, $\mbf w$, in general, is not normalized to a constant number.
In \cite{Korarticle9} we have constructed an example of such a dequantizer 
for the spin $1/2$, for which only the third and the fourth components of $\mbf w$
are related by the normalization condition $w_3+w_4=1$.

The absence of a constant normalization of the tomographic vector 
leads to inconveniences in numerical calculations
and limits the range of use of this approach in practical applications.
Therefore, the question of finding of the projecting states, 
which ensure the fulfillment of the equality
\be			\label{Normw}
\sum_{j=1}^{(2s+1)^2}w_j=\mathrm{const},
\ee
is topical.

The aim of this work is the construction of normalized and non-redundant vector tomographic 
portraits of spin states.

The paper is organized as follows. 
In Sections \ref{Art15Section2} and \ref{Art15Section3} 
the normalized four-component dequantizer vector for the spin $1/2$ without redundancy
is found in general case, the corresponding quantizer vector is calculated,
and their properties are investigated.
In Section \ref{Art15Section4} a graphic geometric interpretation 
of the choice of parameters for finding of realizations of such dequantizers is presented,
and two examples of dequantizers and quantizers are given, 
whose components are pure and mixed states respectively.
In Section \ref{Art15Section5} the procedure for finding the quantizers and dequantizers
is generalized for normalized and non-redundant vector tomographic portraits 
of spin $s=(2^N-1)/2$ states, where $N$ is a natural number.
The conclusion and prospects are presented in \ref{Art15Section6}.

\section{\label{Art15Section2}Normalized non-redundant dequantizer and its properties}
From (\ref{sumw}) it is obvious that for the fulfillment of (\ref{Normw})
for any normalized state $\hat\rho$ it is necessary and sufficient 
that the relation 
$\sum_{j=1}^{(2s+1)^2}\hat{\mathcal{U}}_j=\mathrm{const}\times\hat E$
must be satisfied, where $\hat E$ is the unit $(2s+1)\times(2s+1)$-matrix.
Since the projections $\hat{\mathcal{U}}_ j$ are normalized by the condition
$\mathrm{Tr}\,\hat{\mathcal{U}}_j=1$,
then $\sum_j^{(2s+1)^2}\mathcal{U}_{j(kk)}=2s+1$, $k=\overline{1,2s+1}$.
Therefore, in  (\ref{Normw}) we have the equality $\mathrm{const}=2s+1$, 
i.e., the normalization condition for the vector $\mbf w$ 
has the form
\be			\label{normw2poverN}
\sum_{j=1}^{(2s+1)^2}w_j=2s+1,
\ee
and for the components of the matrix vector $\hat{\bm{\mathcal{U}}}$ 
the following relation is fulfilled: 
\be			\label{sumU}
\sum_{j=1}^{(2s+1)^2}\hat{\mathcal{U}}_j=(2s+1)\times\hat E.
\ee
However, if we multiply the matrices $\hat{\mathcal{U}}_j$ by some weight factor,
we can obtain any preassigned $\mathrm{const}$.

We can also construct a weighted tomographic scheme if instead of (\ref{Normw})
we introduce the requirement for the normalization of the vector $\mbf w $
with the set of weights $\eta_k$ as follows:
$\sum_{j=1}^{(2s+1)^2}\eta_jw_j=1$. Then relation (\ref{sumU}) will take the form:
$\sum_{j=1}^{(2s+1)^2}\eta_j\,\hat{\mathcal{U}}_j=\hat E$.
But in the present paper we restrict ourselves to the case (\ref{normw2poverN}).

Let us find matrices $\hat{\mathcal{U}}_k$ satisfying (\ref{sumU}) for spin $s=1/2$
and explore their properties.

The real 4-component vector $\mbf w$ represents a state if and only if
the density matrix being received from (\ref{invmap}) is Hermitian, non-negative,
and normalized. So, the following relations must be valid:
\be			\label{relationsnonneg1}
\sum_{j=1}^4\mathcal{D}_{j(11)}w_j>0,~~~~
\sum_{j=1}^4\mathcal{D}_{j(22)}w_j>0
\ee
(one of these two sums can also be equal zero), and
\be			\label{relationsnonneg2}
\left(\sum_{i=1}^4\mathcal{D}_{i(11)}w_i\right)\times
\left(\sum_{k=1}^4\mathcal{D}_{k(22)}w_k\right)
-\left|\sum_{l=1}^4\mathcal{D}_{l(12)}w_l\right|^2\geq 0,
\ee
\be			\label{relationnorm}
\sum_{k=1}^4\mathrm{Tr}\{\hat{\mathcal{D}}_k\}w_k=1.
\ee
Hermiticity of $\hat\rho$ is automatically provided owing to the hermiticity 
of $\{\hat{\mathcal{D}}_j\}_{j=\overline{1,4}}$.

If we choose the unit vector $\mbf e_k=(\alpha_k,\beta_k,\gamma_k)$ whose components
are normalized as
\be			\label{normek}
|\mbf e_k|=\alpha_k^2+\beta_k^2+\gamma_k^2=1,
\ee
then the wave vector of the state with the spin projection along $\mbf e_k$,
reliably equal to $1/2$, is found from equation $\mbf e_k\hat{\mbf s}\psi_k=\psi_k/2$,
where $\hat{\mbf s}$ is the spin operator. With the accuracy up to a phase factor we have
\be			\label{VF} 
\psi_k=\frac{1}{\sqrt2}\left(
\begin{array}{c}
\sqrt{\gamma_k+1}  \\[2mm]
\ds{\frac{\alpha_k+i\beta_k}{\sqrt{\gamma_k+1}}}
\end{array}
\right).
\ee
Taking the matrix product $\psi_k\psi_k^\dagger$ we find the projector
corresponding to this state
\be			\label{priectabg}
\hat{\mathcal{U}}_k=\psi_k\psi_k^\dagger=
\frac{1}{2}\left[
\begin{array}{cc}
\gamma_k+1 & \alpha_k-i\beta_k \\
\alpha_k+i\beta_k & 1-\gamma_k
\end{array}
\right].
\ee
From the secular equation we find the eigenvalues $u_{k(1)}$, $u_{k(2)}$
and the determinants of these matrices
\be			\label{eigenvalueUk}
u_{k(1,2)}=\frac{1}{2}\pm\frac{1}{2}\sqrt{\alpha_k^2+\beta_k^2+\gamma_k^2},~~~~
\mathrm{det}\,\hat{\mathcal{U}}_k=u_{k(1)}\,u_{k(2)}=
\left(1-\alpha_k^2-\beta_k^2-\gamma_k^2\right)/4.
\ee
By direct calculation we also find $\mathrm{Tr}\{\hat{\mathcal{U}}_k\hat{\mathcal{U}}_k\}$ 
and $\mathrm{Tr}\{\hat{\mathcal{U}}_j\hat{\mathcal{U}}_k\}$
\bea
\mathrm{Tr}\{\hat{\mathcal{U}}_k\hat{\mathcal{U}}_k\}&=&u_{k(1)}^2+u_{k(2)}^2=
\left(1+\alpha_k^2+\beta_k^2+\gamma_k^2\right)/2,
			\label{TrUUk} \\[3mm]
\mathrm{Tr}\{\hat{\mathcal{U}}_j\hat{\mathcal{U}}_k\}&=&
\left(1+\alpha_j\alpha_k+\beta_j\beta_k+\gamma_j\gamma_k\right)/2.
			\label{TrUjUk}
\eea
Since the matrices $\hat{\mathcal{U}}_k$ correspond to pure states (\ref{VF}), 
for which the vectors $\{\mbf e_k\}$ are normalized by condition (\ref{normek}),
then, as it should be,
$u_{k(1)}=0$, $u_{k(2)}=1$, $\mathrm{det}\,\hat{\mathcal{U}}_k=0$, and
$\mathrm{Tr}\{\hat{\mathcal{U}}_k\hat{\mathcal{U}}_k\}=1$.

However, with the help of sets of values $\{(\alpha_k,\beta_k,\gamma_k)\}_{k=\overline{1,4}}$
we can also parameterize the mixed states $\{\hat{\mathcal{U}}_k\}_{k=\overline{1,4}}$.
For this, the matrices $\hat{\mathcal{U}}_k$ must be positive definite, 
i.e., the following inequalities must be satisfied:
\be			\label{mixedCs}
u_{k(1,2)}>0,~~~~\mathcal{U}_{k(11)}>0,~~~~\mathcal{U}_{k(22)}>0,~~~~
\mathrm{det}\,\hat{\mathcal{U}}_k>0,~~~~
\mathrm{Tr}\left\{\hat{\mathcal{U}}_k^2\right\}<1.
\ee
From formulas (\ref{priectabg},\ref{eigenvalueUk},\ref{TrUUk}) it is obvious that
(\ref{mixedCs}) is fulfilled if
\be			\label{nonormek}
\alpha_k^2+\beta_k^2+\gamma_k^2<1.
\ee
Note also that from (\ref{TrUUk}) the estimate 
$\mathrm{Tr}\{\hat{\mathcal{U}}_k\hat{\mathcal{U}}_k\}>1/2$ follows,
and since $\hat{\mathcal{U}}_k$ can be any density matrix, then this inequality
is true for any state $\hat\rho$ of spin $1/2$, i.e.,
\be			\label{TrRho2}
\mathrm{Tr}\{\hat{\rho}^2\}>1/2.
\ee

Further we will specifically indicate whether we use pure or mixed states $\hat{\mathcal{U}}_k$,
i.e., when the equalities (\ref{normek}) or inequalities (\ref{nonormek}) are fulfilled
respectively, and in the absence of such an indication 
we will assume that our reasoning is true in the general case
for both pure and mixed states.

Matrix equation (\ref{sumU}) is obviously equivalent 
to the following vector equation:
\be			\label{SUMek}
\mbf e_1+\mbf e_2+\mbf e_3+\mbf e_4=0.
\ee
The traces of the products of the matrices $\hat{\mathcal{U}}_j\hat{\mathcal{U}}_k$
satisfy some additional
conditions. To derive them, we write down  formula (\ref{sumU}) as
$\hat{\mathcal{U}}_1+\hat{\mathcal{U}}_2=2\hat{E}-\hat{\mathcal{U}}_3-\hat{\mathcal{U}}_4$,
lift the left and right sides of this equality to a square, 
and take the $\mathrm{Tr}\{\cdot\}$ operation.
Since $4\mathrm{Tr}\hat{E}-4\mathrm{Tr}\,\hat{\mathcal{U}}_3-4\mathrm{Tr}\,\hat{\mathcal{U}}_4=0$,
then
\be			\label{TrU1U2}
2\mathrm{Tr}\{\hat{\mathcal{U}}_1\hat{\mathcal{U}}_2\}+\mathrm{Tr}\{\hat{\mathcal{U}}_1\hat{\mathcal{U}}_1\}+
\mathrm{Tr}\{\hat{\mathcal{U}}_2\hat{\mathcal{U}}_2\}=2\mathrm{Tr}\{\hat{\mathcal{U}}_3\hat{\mathcal{U}}_4\}+
\mathrm{Tr}\{\hat{\mathcal{U}}_3\hat{\mathcal{U}}_3\}+\mathrm{Tr}\{\hat{\mathcal{U}}_4\hat{\mathcal{U}}_4\}.
\ee
By replacing the indexes $2\leftrightarrow3$ or $1\leftrightarrow3$
we obtain formulas for the other remaining products:
\be			\label{TrU1U3}
2\mathrm{Tr}\{\hat{\mathcal{U}}_1\hat{\mathcal{U}}_3\}+ \mathrm{Tr}\{\hat{\mathcal{U}}_1\hat{\mathcal{U}}_1\}+
 \mathrm{Tr}\{\hat{\mathcal{U}}_3\hat{\mathcal{U}}_3\}=2\mathrm{Tr}\{\hat{\mathcal{U}}_2\hat{\mathcal{U}}_4\}+
 \mathrm{Tr}\{\hat{\mathcal{U}}_2\hat{\mathcal{U}}_2\}+ \mathrm{Tr}\{\hat{\mathcal{U}}_4\hat{\mathcal{U}}_4\},
\ee
\be			\label{TrU2U3}
2\mathrm{Tr}\{\hat{\mathcal{U}}_2\hat{\mathcal{U}}_3\}+ \mathrm{Tr}\{\hat{\mathcal{U}}_2\hat{\mathcal{U}}_2\}+
 \mathrm{Tr}\{\hat{\mathcal{U}}_3\hat{\mathcal{U}}_3\}=2\mathrm{Tr}\{\hat{\mathcal{U}}_1\hat{\mathcal{U}}_4\}+
 \mathrm{Tr}\{\hat{\mathcal{U}}_1\hat{\mathcal{U}}_1\}+ \mathrm{Tr}\{\hat{\mathcal{U}}_4\hat{\mathcal{U}}_4\}.
\ee
These formulas are true for both pure and mixed normalized projecting states 
$\hat{\mathcal{U}}_k$.
The only condition for their fulfillment is the requirement (\ref{SUMek}),
where each vector $\mbf e_k$ can have its own normalization, less than or equal to $1$.

If we consider only pure or only mixed projectors $\hat{\mathcal{U}}_k$, 
for which $|\mbf e_1|=|\mbf e_2|=|\mbf e_3|=|\mbf e_4|$,
then from (\ref{TrU1U2} -- \ref{TrU2U3}) we get
\be			\label{TrUjUkPure}
\mathrm{Tr}\{\hat{\mathcal{U}}_1\hat{\mathcal{U}}_2=\mathrm{Tr}\{\hat{\mathcal{U}}_3\hat{\mathcal{U}}_4\},~~~~
\mathrm{Tr}\{\hat{\mathcal{U}}_1\hat{\mathcal{U}}_3=\mathrm{Tr}\{\hat{\mathcal{U}}_2\hat{\mathcal{U}}_4\},~~~~
\mathrm{Tr}\{\hat{\mathcal{U}}_2\hat{\mathcal{U}}_3=\mathrm{Tr}\{\hat{\mathcal{U}}_1\hat{\mathcal{U}}_4\}.
\ee

\section{\label{Art15Section3}The quantizer corresponding to the normalized dequantizer}
We define the matrix $\hat{\mathrm{R}}$ of the dequantizer components $\hat{\bm{\mathcal{U}}}$ 
and the matrix $\hat{\mathrm{J}}$ of the quantizer components $\hat{\bm{\mathcal{D}}}$
as follows:
\be			\label{MatrU}
\hat{\mathrm{R}}=\left(
\begin{array}{cccc}
\mathcal{U}_{1(11)} &\mathcal{U}_{1(21)}&\mathcal{U}_{1(12)}&\mathcal{U}_{1(22)} \\
\mathcal{U}_{2(11)} &\mathcal{U}_{2(21)}&\mathcal{U}_{2(12)}&\mathcal{U}_{2(22)} \\
\mathcal{U}_{3(11)} &\mathcal{U}_{3(21)}&\mathcal{U}_{3(12)}&\mathcal{U}_{3(22)} \\
\mathcal{U}_{4(11)} &\mathcal{U}_{4(21)}&\mathcal{U}_{4(12)}&\mathcal{U}_{4(22)} 
\end{array}
\right),~~~~
\hat{\mathrm{J}}=\left(
\begin{array}{cccc}
\mathcal{D}_{1(11)} &\mathcal{D}_{2(11)}&\mathcal{D}_{3(11)}&\mathcal{D}_{4(11)} \\
\mathcal{D}_{1(12)} &\mathcal{D}_{2(12)}&\mathcal{D}_{3(12)}&\mathcal{D}_{4(12)} \\
\mathcal{D}_{1(21)} &\mathcal{D}_{2(21)}&\mathcal{D}_{3(21)}&\mathcal{D}_{4(21)} \\
\mathcal{D}_{1(22)} &\mathcal{D}_{2(22)}&\mathcal{D}_{3(22)}&\mathcal{D}_{4(22)} 
\end{array}
\right).
\ee
Then relations (\ref{ortcondition}) obviously take on a simple form
$\hat{\mathrm{R}}\times\hat{\mathrm{J}}=\hat1$, where $\hat1$ is the unit
\mbox{4$\times$4-matrix,} i.e.,  
the matrices $\hat{\mathrm{R}}$ and $\hat{\mathrm{J}}$ are mutually inverse, 
and $\hat{\mathrm{J}}=\hat{\mathrm{R}}^{-1}$.
For the existence of an inverse matrix, the condition
\mbox{$\mathrm{det}\,\hat{\mathrm{R}}\not= 0$} is necessary.
The calculation of this determinant with allowance for (\ref{SUMek}) yields
\be			\label{noncomplCond}
\mathrm{det}\,\hat{\mathrm{R}}=i\Delta_1\not= 0,~~~~\mathrm{where} 
~~~\Delta_1=\left|
\begin{array}{ccc}
\alpha_2&\beta_2&\gamma_2 \\[-1mm]
\alpha_3&\beta_3&\gamma_3 \\[-1mm]
\alpha_4&\beta_4&\gamma_4
\end{array}
\right|\not= 0.
\ee
Since the numbering of the four vectors $\{\mbf e_k\}_{k=\overline{1,4}}$ is chosen arbitrarily,
then condition  (\ref{noncomplCond}) means that for the invertibility 
of transformation (\ref{defw}), where the components of the dequantizer
$\hat{\bm{\mathcal{U}}}$ are given by formula (\ref{priectabg}),
it is necessary and sufficient that any thriple of these vectors must not be coplanar, i.e.,
\be			\label{noncomplCond2}
\Delta_2=\left|
\begin{array}{ccc}
\alpha_3&\beta_3&\gamma_3 \\[-1mm]
\alpha_4&\beta_4&\gamma_4 \\[-1mm]
\alpha_1&\beta_1&\gamma_1
\end{array}
\right|\not= 0,~~~
\Delta_3=\left|
\begin{array}{ccc}
\alpha_4&\beta_4&\gamma_4 \\[-1mm]
\alpha_1&\beta_1&\gamma_1 \\[-1mm]
\alpha_2&\beta_2&\gamma_2
\end{array}
\right|\not= 0,~~~
\Delta_4=\left|
\begin{array}{ccc}
\alpha_1&\beta_1&\gamma_1 \\[-1mm]
\alpha_2&\beta_2&\gamma_2 \\[-1mm]
\alpha_3&\beta_3&\gamma_3
\end{array}
\right|\not= 0,
\ee
and from (\ref{SUMek}) it follows that $\Delta_1=-\Delta_2=\Delta_3=-\Delta_4$.
Using the Cramer rule and properties of determinants known from the linear algebra,
taking into account (\ref{SUMek})
we find the components of the quantizer $\hat{\bm{\mathcal{D}}}$:
\bea			\label{QuantizerComponents1}
\mathcal{D}_{1(11)}&=&-\frac{1}{4\Delta_1}\left|
\begin{array}{ccc}
1&\alpha_2&\beta_2 \\[-1mm]
1&\alpha_3&\beta_3 \\[-1mm]
1&\alpha_4&\beta_4
\end{array}
\right|+\frac{1}{4},~~~~
\mathcal{D}_{1(22)}=\frac{1}{4\Delta_1}\left|
\begin{array}{ccc}
1&\alpha_2&\beta_2 \\[-1mm]
1&\alpha_3&\beta_3 \\[-1mm]
1&\alpha_4&\beta_4
\end{array}
\right|+\frac{1}{4}, \nonumber \\
\mathcal{D}_{1(12)}&=&-\frac{1}{4\Delta_1}\left(\left|
\begin{array}{ccc}
1&\beta_2&\gamma_2 \\[-1mm]
1&\beta_3&\gamma_3 \\[-1mm]
1&\beta_4&\gamma_4
\end{array}
\right|+i\left|
\begin{array}{ccc}
1&\alpha_2&\gamma_2 \\[-1mm]
1&\alpha_3&\gamma_3 \\[-1mm]
1&\alpha_4&\gamma_4
\end{array}
\right|
\right),~~~~\mathcal{D}_{1(21)}=\left[\mathcal{D}_{1(12)}\right]^*.
\eea
Similar formulas for $\hat{\mathcal{D}}_2$, $\hat{\mathcal{D}}_3$, and $\hat{\mathcal{D}}_4$
are obtained from (\ref{QuantizerComponents1}) by cyclic permutation of indexes $1,2,3,4$
corresponding to the components of the tomographic vector $\mbf w$.

Let us study the properties of the matrices $\{\hat{\mathcal{D}}_k\}_{k=\overline{1,4}}$ obtained.
From (\ref{QuantizerComponents1}) it is clear that these matrices are
Hermitian and normalized by the condition
\be			\label{normD}
\mathrm{Tr}\,\hat{\mathcal{D}}_k =1/2,~~~~k=\overline{1,4}.
\ee
Adding the matrices $\hat{\mathcal{D}}_k$ after calculations we get
\be			\label{sumD}
\sum_{k=1}^4\hat{\mathcal{D}}_k=\hat E.
\ee
From the secular equation we easily find the eigenvalues $d_{1(1)}$ and $d_{1(2)}$ 
of the matrix $\hat{\mathcal{D}}_1$
\be			\label{eigenvalue}
d_{1(1,2)}=\frac{1}{4}\pm \frac{1}{4\Delta_1}\left(
\left|
\begin{array}{ccc}
1&\alpha_2&\beta_2\\[-1mm]
1&\alpha_3&\beta_3\\[-1mm]
1&\alpha_4&\beta_4
\end{array}
\right|^2
+\left|
\begin{array}{ccc}
1&\beta_2&\gamma_2\\[-1mm]
1&\beta_3&\gamma_3\\[-1mm]
1&\beta_4&\gamma_4
\end{array}
\right|^2
+\left|
\begin{array}{ccc}
1&\alpha_2&\gamma_2\\[-1mm]
1&\alpha_3&\gamma_3\\[-1mm]
1&\alpha_4&\gamma_4
\end{array}
\right|^2
\right)^{1/2}.
\ee
The eigenvalues of the matrices $\hat{\mathcal{D}}_2$, $\hat{\mathcal{D}}_3$, and 
$\hat{\mathcal{D}}_4$  are obtained from (\ref{eigenvalue}) by means of a cyclic 
permutation of the indexes $1,2,3,4$.
Knowing the eigenvalues $d_{k(1,2)}$, we find 
$\mathrm{det}\hat{\mathcal{D}}_k=d_{k(1)}d_{k(2)}$ and
$\mathrm{Tr}\{\hat{\mathcal{D}}_k\hat{\mathcal{D}}_k\}=d_{k(1)}^2+d_{k(2)}^2$.

The matrices $\hat{\mathcal{D}}_k$ are negative definite; one of the eigenvalues
$d_{k(1,2)}$ is negative, and the other is positive, i.e., 
$\mathrm{det}\hat{\mathcal{D}}_k=d_{k(1)}d_{k(2)}< 0$.

For example, we prove this statement for $\hat{\mathcal{D}}_1$.
Since according to (\ref{QuantizerComponents1}) 
$\hat{\mathcal{D}}_1$ is a Hermitian matrix, then with the aid of 
a unitary transformation we can reduce it to the diagonal form
\be			\label{diagonal}
\hat{\mathfrak{V}}^{-1}\hat{\mathcal{D}}_1\hat{\mathfrak{V}}=
\left(
\begin{array}{cc}
d_{1(1)}&0\\
0&d_{1(2)}
\end{array}
\right),~~~~
\hat{\mathcal{D}}_1=
\hat{\mathfrak{V}}\left(
\begin{array}{cc}
d_{1(1)}&0\\
0&d_{1(2)}
\end{array}
\right)\hat{\mathfrak{V}}^{-1},
\ee
where $\hat{\mathfrak{V}}$ is a unitary matrix, whose columns 
are eigenvectors of the matrix $\hat{\mathcal{D}}_1$.
Substituting (\ref{diagonal}) into (\ref{ortcondition}) and using 
the properties of the $\mathrm{Tr}\{\cdot\}$ operation we get
\be			\label{Trconv}
\mathrm{Tr}\left\{\hat{\mathcal{U}}_k
\hat{\mathcal{D}}_1\right\}=
\mathrm{Tr}\left\{
\hat{\mathcal{U}}'_k
\left(
\begin{array}{cc}
d_{1(1)}&0\\
0&d_{1(2)}
\end{array}
\right)
\right\}=
\mathcal{U}_{k(11)}'d_{1(1)}+
\mathcal{U}_{k(22)}'d_{1(2)}=\delta_{1k}\,,
\ee
where the notation 
$\hat{\mathcal{U}}'_k=\hat{\mathfrak{V}}^{-1}\hat{\mathcal{U}}_k\hat{\mathfrak{V}}$
was introduced. 
Since $\hat{\mathcal{U}}_k$ is a non-negative definite normalized matrix, then 
$\hat{\mathcal{U}}'_k$ are also non-negative definite and normalized. Therefore
$\mathcal{U}_{k(11)}'\geq 0$, $\mathcal{U}_{k(22)}'\geq 0$, and
$\mathcal{U}_{k(11)}'+\mathcal{U}_{k(22)}'=1.$

From (\ref{Trconv}) we have four equations:
\be			\label{foureq}
\begin{array}{ll}
\mathcal{U}_{1(11)}'d_{1(1)}+\mathcal{U}_{1(22)}'d_{1(2)}=1, & 
\mathcal{U}_{2(11)}'d_{1(1)}+\mathcal{U}_{2(22)}'d_{1(2)}=0,\\[3mm]
\mathcal{U}_{3(11)}'d_{1(1)}+\mathcal{U}_{3(22)}'d_{1(2)}=0, & 
\mathcal{U}_{4(11)}'d_{1(1)}+\mathcal{U}_{4(22)}'d_{1(2)}=0.
\end{array}
\ee
To satisfy these equations it is necessary that one of the eigenvalues of the matrix
$\hat{\mathcal{D}}_1$ be positive and the other be negative. Thus, 
$\hat{\mathcal{D}}_1$ is negative definite. Similarly, 
negative definiteness is proved for the matrices $\hat{\mathcal{D}}_2$,
$\hat{\mathcal{D}}_3$, and $\hat{\mathcal{D}}_4$.

We also point out that since (\ref{sumD}) is analogous to the equality (\ref{sumU})
up to a coefficient of 2
and $\mathrm{Tr}\,\hat{\mathcal{U}}=2\,\mathrm{Tr}\hat{\mathcal{D}}$, 
then the traces of the products $\hat{\mathcal{D}}_j\hat{\mathcal{D}}_k$
satisfy the same equalities (\ref{TrU1U2}--\ref{TrU2U3})
as the traces of the products $\hat{\mathcal{U}}_j\hat{\mathcal{U}}_k$.

\section{\label{Art15Section4}Examples of dequantizers and quantizers}
First of all, we indicate that relations (\ref{normek}) and/or (\ref{nonormek}),
(\ref{SUMek}), and (\ref{noncomplCond}) or (\ref{noncomplCond2}) 
for the components of the vectors $\mbf e_k$, which determine the conditions 
for the existence of a reversible and normalized dequantizer, 
admit a simple geometric interpretation.

\begin{figure*}
\begin{center}
\begin{minipage}{0.8\linewidth}
\center{\includegraphics[width=0.8\linewidth,height=0.25\linewidth]{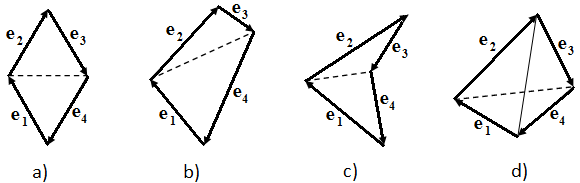}
\caption{\label{figure1} a), b), c) examples of initial quadrangles on a plane;
d) a pyramid obtained by bending a quadrangle along a diagonal indicated by a dashed line.}}
\end{minipage}
\end{center}
\end{figure*}

Let us construct an arbitrary quadrangle on the plane with sides less than or equal $1$.
Then let us choose the direction of the bypass of this quadrangle and determine 
at each side the beginning and the end in accordance with this direction.
If you bend such a quadrangle along any of the diagonals, you get a triangular pyramid.
Figure \ref{figure1} shows examples of quadrangles, 
bending of which along the diagonals indicated by the dashed lines  
yields a pyramid.
Carrying out the turns of this pyramid in space, we can orient it arbitrarily.

The four edges of this pyramid corresponding to the directed sides of the original 
quadrangle form a quadruple of vectors $\{\mbf e_k\}_{k=\overline{1,4}}$
satisfying (\ref{SUMek}). 
If some of these edges has a length equal to one, then it corresponds to a pure state
$\hat{\mathcal{U}}_k$. The edges with a lengthes less than one correspond to mixed states.
According to (\ref{noncomplCond}) or (\ref{noncomplCond2})
the volume of our pyramid should not be zero, i.e., the pyramid should not be degenerate.

Since there are an infinite number of such pyramids, then there are infinite number 
of possible normalized dequantizers, and
having such a clear geometric interpretation, it is not difficult to find examples of them.

\textbf{Example 1. Pure states.} 
Choose the vectors $\{\mbf e_k\}$ as follows: $\mbf e_1=(0,4,3)/5$, $\mbf e_2=(4,0,-3)/5$,
$\mbf e_3=(0,-4,3)/5$, $\mbf e_4=(-4,0,-3)/5$.
This is the case of pure states $\{\hat{\mathcal{U}}_k\}_{k=\overline{1,4}}$
because all four vectors are normalized to 1.
With the help of (\ref{priectabg}) and (\ref{QuantizerComponents1}) we find dequantizer
$\hat{\bm{\mathcal{U}}}^{(1)}$ and quantizer $\hat{\bm{\mathcal{D}}}^{(1)}$ respectively
\bea
&&
\hat{\bm{\mathcal{U}}}^{(1)}=\frac{1}{5}\left(
\left[\begin{array}{cc} 4 & -2i \\2i & 1\end{array}\right],
\left[\begin{array}{cc} 1 & 2  \\ 2  & 4\end{array}\right],
\left[\begin{array}{cc} 4 & 2i \\-2i  & 1\end{array}\right],
\left[\begin{array}{cc} 1 & -2  \\-2 & 4\end{array}\right]
\right), \nonumber \\[3mm]
&&
\hat{\bm{\mathcal{D}}}^{(1)}=\frac{1}{2}\left(
\left[\begin{array}{cc} 4/3&-5i/4 \\5i/4&-1/3\end{array}\right],
\left[\begin{array}{cc} -1/3&5/4 \\5/4&4/3\end{array}\right],
\left[\begin{array}{cc} 4/3&5i/4 \\-5i/4&-1/3\end{array}\right],
\left[\begin{array}{cc} -1/3&-5/4 \\-5/4&4/3\end{array}\right]
\right). \nonumber
\eea

\textbf{Example 2. Mixed states.}
We take the four vectors
$\mbf e_1=(0,-2,1)/3$, $\mbf e_2=(2,0,-1)/3$,
$\mbf e_3=(0,2,1)/3$, $\mbf e_4=(-2,0,-1)/3$
normalized to the same number $\sqrt5/3$,
which is less then 1.
Then, after calculations, we obtain the dequantizer $\hat{\bm{\mathcal{U}}}^{(2)}$
with the components that are mixed states,
and the corresponding quantizer $\hat{\bm{\mathcal{D}}}^{(2)}$
\bea
&&
\hat{\bm{\mathcal{U}}}^{(2)}=\frac{1}{3}\left(
\left[\begin{array}{cc} 2 & i \\-i & 1\end{array}\right],
\left[\begin{array}{cc} 1 & 1  \\ 1  & 2\end{array}\right],
\left[\begin{array}{cc} 2 & -i \\ i  & 1\end{array}\right],
\left[\begin{array}{cc} 1 & -1  \\-1 & 2\end{array}\right]
\right), \nonumber \\[3mm]
&&
\hat{\bm{\mathcal{D}}}^{(2)}=\frac{1}{2}\left(
\left[\begin{array}{cc} 2&3i/2 \\-3i/2&-1\end{array}\right],
\left[\begin{array}{cc} -1&3/2 \\3/2&2 \end{array}\right],
\left[\begin{array}{cc} 2&-3i/2 \\3i/2&-1\end{array}\right],
\left[\begin{array}{cc} -1&-3/2 \\-3/2&2\end{array}\right]
\right). \nonumber
\eea

\section{\label{Art15Section5}Generalization to the case of spin $s=(2^N-1)/2$}
The problem of finding of normalized non-redundant dequantizers and corresponding quantizers 
for large values of spins in general case turns out to be nontrivial.

At the same time, for single spins $s=(2^N-1)/2$, where $N$ is a natural number,
the application of the well known technique is possible.
The point is that the $(2^N)\times(2^N)$-density matrices with the components $\rho_{k,l}$ 
for such a quantum system can be treated as the $\underbrace{2\times...\times2}_{2N}$-density 
matrices with the components $\rho_{k_1k_2...k_N,l_1l_2...l_N}=\rho_{k,l}$
for the system of $N$ spins 1/2 using some one-to-one correspondence $\mathfrak{g}$
of sets of indexes
\be			\label{corrInd}
\left\{\,k\, \left|\,k=\overline{1,(2s+1)}\right.\right\}
\overset{\mathfrak{g}}{\underset{\mathfrak{g}^{-1}}{\rightleftarrows}}
\Big\{\,(k_1,k_2,...,k_N)\, \Big|\,k_1,k_2,...,k_N=\overline{1,2}\Big\}.
\ee
Note that this approach seems to be fruitful for the realization of quantum computations,
whose algorithms can be modeled by the evolution of systems of qubits.

The components of the dequantizer $\hat{\bm{\mathfrak{U}}}$ for 
$\rho_{k_1k_2...k_N,l_1l_2...l_N}$
can be introduced as direct products of components of dequantizers
$\hat{\bm{\mathcal{U}}}^{(1)},\hat{\bm{\mathcal{U}}}^{(2)},...,\hat{\bm{\mathcal{U}}}^{(N)}$ 
for spin 1/2, and these dequantizers can both be the same,
$\hat{\bm{\mathcal{U}}}^{(1)}=\hat{\bm{\mathcal{U}}}^{(2)}=...=\hat{\bm{\mathcal{U}}}^{(N)}$,
and be different,
\be			\label{DequantN}
\hat{\mathfrak{U}}_{j_1j_2...j_N}=\hat{\mathcal{U}}_{j_1}^{(1)}\otimes\hat{\mathcal{U}}_{j_2}^{(2)}\otimes
...\otimes\hat{\mathcal{U}}_{j_N}^{(N)},~~~~
j_1,j_2,...,j_N=\overline{1,4}\,.
\ee
The corresponding quantizer $\hat{\bm{\mathfrak{D}}}$ will have the following components:
\be			\label{QuantN}
\hat{\mathfrak{D}}_{j_1j_2...j_N}=\hat{\mathcal{D}}_{j_1}^{(1)}\otimes\hat{\mathcal{D}}_{j_2}^{(2)}\otimes
...\otimes\hat{\mathcal{D}}_{j_N}^{(N)},
\ee
where to each dequantizer $\hat{\bm{\mathcal{U}}}^{(i)}$ there corresponds 
its own quantizer $\hat{\bm{\mathcal{D}}}^{(i)}$.
The product of the components $\hat{\bm{\mathfrak{U}}}$ and $\hat{\bm{\mathfrak{D}}}$
will be defined as follows:
\be			\label{DquantNtimesQuantN}
\hat{\mathfrak{U}}_{j_1j_2...j_N}\hat{\mathfrak{D}}_{k_1k_2...k_N}=
\left(\hat{\mathcal{U}}_{j_1}\hat{\mathcal{D}}_{k_1}\right)\otimes\left(\hat{\mathcal{U}}_{j_2}\hat{\mathcal{D}}_{k_2}\right)\otimes
...\otimes\left(\hat{\mathcal{U}}_{j_N}\hat{\mathcal{D}}_{k_N}\right),
\ee
from which the orthogonality and completeness conditions immediately follow,
\be			\label{OrtogN}
\mathrm{Tr}\left\{\hat{\mathfrak{U}}_{j_1j_2...j_N}\hat{\mathfrak{D}}_{j_1'j_2'...j_N'} \right\}=
\delta_{j_1j_1'}\delta_{j_2j_2'}...\delta_{j_Nj_N'}~,
\ee
\be			\label{PolnN}
\sum_{j_1,j_2,...,j_N=1}^4
\mathfrak{U}_{j_1(k_1l_1)j_2(k_2l_2)...j_N(k_Nl_N)}
\mathfrak{D}_{j_1(k_1'l_1')j_2(k_2'l_2')...j_N(k_N'l_N')}=
\delta_{k_1k_1'}\delta_{l_1l_1'}...\delta_{k_Nk_N'}\delta_{l_Nl_N'}.
\ee
Using the reverse renaming of indexes
$(k_1,k_2,...,k_N) \overset{\mathfrak{g^{-1}}}{\longrightarrow} k$, 
$(l_1,l_2,...,l_N)\overset{\mathfrak{g^{-1}}}{\rightarrow}l$
with the help of (\ref{corrInd}) and the re-designation of indexes
$(j_1,j_2,...,j_N)\overset{\mathfrak{f}}{\rightarrow}j$
with the help of a some one-to-one correspondence $\mathfrak{f}$
\be			\label{corrIndj}
\Big\{\,(j_1,j_2,...,j_N)\, \Big|\,j_1,j_2,...,j_N=\overline{1,4}\Big\}
\overset{\mathfrak{f}}{\underset{\mathfrak{f}^{-1}}{\rightleftarrows}}
\left\{\,j\, \left|\,j=\overline{1,(2s+1)^2}\right.\right\},
\ee
we can bring $\hat{\bm{\mathfrak{U}}}$ and $\hat{\bm{\mathfrak{D}}}$
to the form, in which they will represent $(2s+1)^2$-component vectors 
with components of $(2s+1)\times(2s+1)$-matrices
\be			\label{VectDequantNorg}
\mathfrak{U}_{j(kl)}=\mathfrak{U}_{j_1(k_1l_1)j_2(k_2l_2)...j_N(k_Nl_N)},
\ee
\be			\label{VectQuantNorg}
\mathfrak{D}_{j(kl)}=\mathfrak{D}_{j_1(k_1l_1)j_2(k_2l_2)...j_N(k_Nl_N)},
\ee
where $j=\overline{1,(2s+1)^2}$ is the index of the component of the vector $\hat{\bm{\mathfrak{U}}}$
or $\hat{\bm{\mathfrak{D}}}$, and $(kl)$ are the spin indexes of the $(2s+1)\times(2s+1)$-matrices, 
$k,l=\overline{1,(2s+1)}$.

If in (\ref{DequantN}) we now choose the dequantizers
$\hat{\bm{\mathcal{U}}}^{(1)}$, $\hat{\bm{\mathcal{U}}}^{(2)}$, ..., 
$\hat{\bm{\mathcal{U}}}^{(N)}$ so that their components satisfy (\ref{sumU}),
then $\hat{\bm{\mathfrak{U}}}$ and $\hat{\bm{\mathfrak{D}}}$
will automatically be normalized as follows:
\be			\label{normDquantQquantN}
\sum_{j=1}^{(2s+1)^2} \mathfrak{U}_{j(kl)}=
2^N\delta_{kl},~~~~
\sum_{j=1}^{(2s+1)^2} \mathfrak{D}_{j(kl)}=
\delta_{kl},
\ee
\be			\label{TrUjN_TrDjN}
\mathrm{Tr}\,\hat{\mathfrak{U}}_j=1,~~~~
\mathrm{Tr}\,\hat{\mathfrak{D}}_j=1/2^N,~~~~
j=\overline{1,(2s+1)^2}.
\ee
Thus, we have constructed the normalized dequantizer $\hat{\bm{\mathfrak{U}}}$ 
and quantizer $\hat{\bm{\mathfrak{D}}}$ satisfying orthogonality and completeness 
conditions (\ref{ortcondition}) for density matrices of the order of $2^N\times2^N$. 
By means of conversion (\ref{defw}) with use of $\hat{\bm{\mathfrak{U}}}$ 
such density matrices are transformed to
$4^N$-component non-redundant tomographic vectors $\mbf w$
with nonnegative components normalized by the condition (\ref{normw2poverN}),
where $2s+1=2^N$, and the inverse transformation are given 
by (\ref{invmap}) with use of $\hat{\bm{\mathfrak{D}}}$.

\section{\label{Art15Section6}Conclusion}
In conclusion, we point out the main results of the paper.
The positive four-component non-redundant normalized vector  tomographic portrait 
fully describing the  states of spin-$1/2$ quantum particles was introduced 
and it was shown that such a portrait is defined by the thriple 
of non-complanar vectors with the lengthes equal or less then unity.

The corresponding dequantizer and quantizer for spin $1/2$  were found in
general case and their properties were explored.
In particular, it was shown that the vector-quantizer also turns out to be normalized.

A graphic geometric interpretation of the choice of parameters
for finding of numerical realizations of four-component normalized
vectors-dequantizers for the spin $1/2$ was given
and two examples of such dequantizers and quantizers were presented,
whose components are pure and mixed states respectively.

It was also done the generalization of the procedure for finding
of normalized and non-redundant dequantizers and quantizers for spin $s=(2^N-1)/2$,
where $N$ is a natural number.

The normalized tomographic portrait proposed in this paper is useful for
constructing of a set of tomographic schemes and also for realizing
of quantum calculations whose algorithms can be modeled by evolution processes
of systems of qubits and qudits.


\end{document}